\documentclass[12pt, draftclsnofoot, onecolumn]{IEEEtran}

\usepackage{cite}

\usepackage{graphicx}
\usepackage{amsmath}

\usepackage{algorithmic}
\usepackage{color}
\usepackage[table,xcdraw]{xcolor}
\usepackage{multirow}
\usepackage{soul}
\ifCLASSOPTIONcompsoc
  \usepackage[caption=false,font=normalsize,labelfont=sf,textfont=sf]{subfig}
\else
  \usepackage[caption=false,font=footnotesize]{subfig}
\fi

\newtheorem{lemma}{Lemma}
\newtheorem{remark}{Remark}

\begin{document}
\title{Joint Power Minimization Over Multi-Carrier Two-Way Relay Networks}
\author{Zhi~Chen~\IEEEmembership{Member,~IEEE}, Pin-Han~Ho~\IEEEmembership{Senior Member,~IEEE} and Limei Peng~\IEEEmembership{Member,~IEEE}
\thanks{Z. Chen and P. Ho are with the Department of Electrical and Computer
Engineering, University of Waterloo, Waterloo, Ontario, Canada, N2L3G1.
Emails: chenzhi2223@gmail.com; p4ho@uwaterloo.ca.

Limei Peng is with the Department of Industrial and Information System Engineering, Ajou University, , Suwon 443-749, South Korea.}}

\maketitle

\maketitle
\baselineskip 24pt
\begin{abstract}\\
\baselineskip=18pt
The study considers a three-node, two-way relaying network (TWRN) over a multi-carrier system, aiming to minimize the total power consumption of all the transmit and receiver activities. By employing digital network coding (DNC) and physical-layer network coding (PNC), respectively, as well as a novel hybrid PNC/DNC switching scheme, the total transmission power of the considered multi-carrier TWRN system is firstly analyzed; and the derived analytical expressions are then used to formulate a set of nonconvex optimization problems. We will show how those nonconvex functions are convexified for better computational tracbility. It is observed in the numerical results that, the PNC scheme generally consumes less power than DNC but becomes worse in the very low SNR regime. The proposed hybrid PNC/DNC switching scheme that takes advantage of both, is shown to outperform in all SNR regimes.

\end{abstract}

\IEEEpeerreviewmaketitle

\section{Introduction}
Network coding has been proved as an efficient approach in improving network performance. Two-way relay networks (TWRNs) \cite{liu2008network}-\cite{wilson2010joint} is one of the network coding applications, where the relay node can transmit a combined message to both source nodes for decoding. Since each source node has prior knowledge of its originated message, it can subtract this message and obtain the intended message from the other source, thereby improving the spectrum efficiency.

In the literature, several network coding strategies were developed for TWRNs to achieve a larger rate region, such as digital network coding (DNC), physical-layer network coding (PNC), amplify-network-coding (ANC). In DNC, the relay node is required to decode both messages and then combine them together for forwarding along the downlink \cite{liu2008network}-\cite{Berry2}, where the rate region is firstly discussed in \cite{liu2008network}.

In PNC, whereas, the relay node only decodes a function message of the two source messages and then forwards them along the downlink \cite{Liew2006}-\cite{popovski2007physical}, which is expected to yield a higher rate region. A benchmark work on the achievable rate region of PNC for symmetric traffic in TWRNs was presented in \cite{wilson2010joint}. In ANC, the relay node only amplifies the superimposed signals received from the uplink and broadcasts it over the downlink \cite{zhang2009optimal}-\cite{ho2008two}.
Optimal transmit power allocation under ANC for symmetric rate requirement was investigated in \cite{pischella2011optimal} without jointly optimizing the time resources; whereas, time splitting for ANC based TWRNs under a given rate requirement was studied in \cite{hausl2012resource}. In \cite{ChenTVT'14}, the symmetric end-to-end throughput on both sides over fading channels was considered, where the opportunistic three-slot DNC and PNC was considered.
\cite{Chentwo-way'16} investigated power efficiency in several DNC related strategies over Gaussian channels. In \cite{Simmons'15}, the authors considered with optimal power allocation over two-way relaying with AF mode and peak power constraints.

To the best of our survey, the asymmetric traffic scenario in TWRNs has been insufficiently studied. Further, there has never been any research/study that takes the receive-side power consumption into consideration in their analytical formulations and optimization frameworks. Note that the TWRN systems are envisioned to possibly serve in the future IoT and sensor networks, where all the nodes, including the sources and the relay, are battery-powered and their power consumptions should be jointly considered and taken into the optimization framework.


Rather than ANC whose time resource can only be evenly allocated to the uplink and downlink transmissions, this work targets at multi-carrier TWRNs over static fading channels by using a network coding scheme based on DNC, PNC or a novel hybrid DNC/PCN scheme that intelligently switches between the two, subject to arbitrary rate requirements from both sides. Specifically, we are committed to provide an optimization framework for total power consumption, where both the time and power resources at the sources and relays are jointly determined. The contributions of the paper are hence summarized as follows:
\begin{itemize}

\item
We formulated the problems of total transmit and receive power minimization in a multi-carrier two-way relay network, subject to arbitrary rate requirements of both sources. The three network coding strategies, namely, DNC, PNC, and a hybrid scheme of DNC/PNC, are taken in our analysis.

\item The formulated optimization problems originally presented as nonconvex are convexified for better computational tractability, where close-form global optimal solutions are derived via the Karush-Kuhn-Tucker (KKT) conditions.


\item Based on the extensive simulation results, we conclude that PNC performs well under mid to high rates, while DNC takes advantages under low rate pair requirements. The proposed hybrid PNC/DNC switching scheme, as expected, outperforms both of the schemes over all the SNR regimes due to its design premises that exploits the best of the two.

\end{itemize}


This paper is organized as follows. Section II describes the system model and the network coding strategies considered in this work. In Sections III and IV, DNC and PNC are discussed, respectively, and the associated optimization problems are formulated and analyzed. The proposed hybrid PNC/DNC switching scheme as well as the proposed iterative algorithm is presented in Section V, which exploits the best of the two schemes. The convergence as well as complexity of the proposed iterative algorithm is also discussed. The numerical results are presented in Section VI, and the paper is concluded in Section VII.

\section{System Description}
Consider a three-node, two-way relaying network (TWRN) consisting of two sources $S_1$, $S_2$ and one relay $R$ over $N$ orthogonal subcarriers (SCs) over Gaussian channels. The two sources intend to exchange messages through the relay node $R$ via the $N$ subchannels, where a direct link between them is not feasible. Our goal is to minimize the total power consumption of the TWRN over the $N$ independent orthogonal sub-channels, subject to rate requirements from both sides, denoted as $\lambda_{1}$ and $\lambda_2$ for $S_1$ and $S_2$, respectively.

We assume $\lambda_1 \leq \lambda_2$ in the following analysis due to the random labeling of the source nodes. Let $g_{1r}(n)$, $g_{2r}(n)$, $g_{r1}(n)$ and $g_{r2}(n)$ denote the channel gains for the links $S_1$-$R$, $S_2$-$R$, $R$-$S_1$ and $R$-$S_2$, respectively, over the $n$th carrier. Assume the noise of each channel to be an additive white Gaussian distributed random variable with zero mean and unity variance. A peak power constraint, i.e., $P_{peak}$, is imposed on all transmit nodes. The receiver-side power consumption at each node is assume to consume a constant $P_{rev}$ power level in message decoding when it is in the "on" period in receiving messages and zero otherwise. Further, we assume that all nodes work in a half-duplex mode, i.e., at any moment they can either transmit or receive messages, but not both.

\begin{figure}[t]
   \centering
   \includegraphics[width = 12cm]{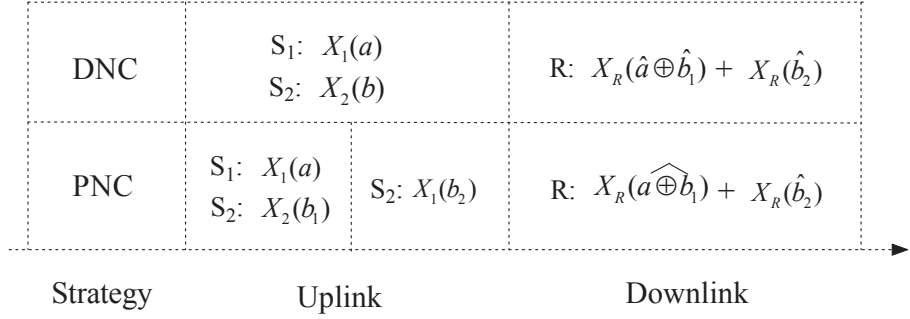}
   \caption{Illustration of the considered DNC and PNC schemes, where message $a$ and message $b_1$ are of the length. } \label{fig:model}
   \end{figure}

Fig. \ref{fig:model} demonstrates the TWRN transmission under DNC and PNC. Here $a$ and $b$ denotes the message sent by $S_1$ and $S_2$, respectively, where we assume $b=b_1+b_2$ and $b_1$ and $a$ are of the same length. $X_i(\cdot)$ denotes the codeword of the message ($\cdot$). With DNC, the relay decodes the two individual source messages on the uplink, i.e., both $\hat{a}$ and $\hat{b}$. With PNC, the relay node decodes the function message $\widehat{a+b_1}$ and the remaining bits $b_2$.

The three network coding schemes, namely DNC, PNC, and hybrid DNC/PNC, will be described in detailed in the following sections, along with the formulated optimization problems in minimizing the total power consumption for the considered multi-carrier TWRNs.

\section{DNC}
With DNC, we allow for a multi-access uplink and superposition of network-coded message at the relay node, and the remaining bits of the message to be with a larger size on the downlink, in order to minimize the total power consumption while guaranteeing the specific rate requirement. The two transmission phases of DNC are defined as follows.
\begin{itemize}
\item {\em Uplink Transmission Phase:} $S_1$ and $S_2$ can simultaneously transmit to $R$ at rates $R^{\mathrm{DNC}}_{11}(n)$ and $R^{\mathrm{DNC}}_{12}(n)$ in the multi-access uplink over the $n$th carrier, respectively, with the required source powers denoted as $P^{\mathrm{DNC}}_{11}(n)$ and $P^{\mathrm{DNC}}_{12}(n)$.
\item {\em Downlink Transmission Phase:} $R$ can broadcast to $S_1$ and $S_2$ at rate pair ($R^{\mathrm{DNC}}_{21}(n), R^{\mathrm{DNC}}_{22}(n)$) in downlink over the $n$th carrier. $R^{\mathrm{DNC}}_{22}(n)$ is the network coded data and required by both users, and $R^{\mathrm{DNC}}_{21}(n)$ is only for $S_1$ that further needs the remaining bits of the larger message from $S_2$, due to the assumption that $\lambda_1 \leq \lambda_2$. The required powers for supporting $R^{\mathrm{DNC}}_{21}(n)$ and $R^{\mathrm{DNC}}_{22}(n)$ are denoted as $P^{\mathrm{DNC}}_{21}(n)$ and $P^{\mathrm{DNC}}_{22}(n)$, respectively. \footnote{Note that the network-coded message length is determined by the minimum of $\lambda_1$ and $\lambda_2$, i.e., $\lambda_1$ from the assumption that $\lambda_1 \leq \lambda_2$.}
\end{itemize}

For convenience, we denote $P^{\mathrm{DNC}}_1(n)$ and $P^{\mathrm{DNC}}_2(n)$ as the total transmit power on the uplink and the downlink over carrier $n$, respectively. According to the multi-access achievable rate region in \cite{cover1991elements} and \cite{David}, if $g_{1r}(n)>g_{2r}(n)$, the transmit rate of each user on the $n$th carrier is given by
\begin{align}
R^{\mathrm{DNC}}_{11}(n) &= \log_2\left(1+\frac{P^{\mathrm{DNC}}_{11}(n)g_{1r}(n)}{1+P^{\mathrm{DNC}}_{12}(n)g_{2r}(n)}\right)\label{eqn:MA1} \\
R^{\mathrm{DNC}}_{12}(n) &= \log_2\left(1+P^{\mathrm{DNC}}_{12}(n)g_{2r}(n)\right)  \label{eqn:MA2}
\end{align}
Note that the total power consumption is determined by the decoding order and the channel gains \cite{David}, which can be manipulated for achieving the best system performance. Since the messages with higher channel gains can tolerate more interferences than those with the weaker channel, the weaker-channel messages should be decoded later in order to achieve higher power efficiency.

From (\ref{eqn:MA1}) and (\ref{eqn:MA2}), the optimal transmit powers of $S_1$ and $S_2$ for a given channel gain pair in carrier $n$ can be readily obtained and hence
Then, the total power consumption on the uplink for each carrier is given by,
\begin{align}
P^{\mathrm{DNC}}_{1}(n)=
&\frac{2^{R^{\mathrm{DNC}}_{11}(n)+R^{\mathrm{DNC}}_{12}(n)}}{g_{1r}(n)}- \frac{1}{g_{2r}(n)}\\
&+2^{R^{\mathrm{DNC}}_{12}(n)}\left( \frac{1}{g_{2r}(n)}
-\frac{1}{g_{1r}(n)}  \right). \label{Eqn:rate13}
\end{align}
It can be observed that the total uplink power consumption for each carrier in (\ref{Eqn:rate13}) is a convex function for the associated uplink rates from both source nodes.

In downlink, the network coded data is required by both sources, hence is determined by the minimum of the channel gains. On the other hand, the remaining bits are only required by $S_1$ and known a priori at $S_2$. In the case that $g_{r1}(n)>g_{r2}(n)$, we hence have that
\begin{align}
P^{\mathrm{DNC}}_{21}(n)= & \frac{2^{R^{\mathrm{DNC}}_{21}(n)}-1}{g_{r1}(n)} \label{eq:dnc_uplink_1}\\
P^{\mathrm{DNC}}_{22}(n)
=&\max \Biggl( \frac{2^{R^{\mathrm{DNC}}_{22}(n)}-1}{g_{r2}(n)} , \label{eq:dnc_uplink_1} \\ &\frac{2^{R^{\mathrm{DNC}}_{21}(n)}\left(2^{R^{\mathrm{DNC}}_{22}(n)}-1\right)}{g_{r1}(n)} \Biggr). \nonumber
\end{align}
Note that $P^{\mathrm{\mathrm{DNC}}}_{21}(n)$ is the minimum power consumption required by decoding the remaining bits of the larger message at $S_1$ after decoding the network-coded message. For $P^{\mathrm{\mathrm{DNC}}}_{22}(n)$ of the network coded message, the first term is the minimum power consumption at $S_2$ due to its ability to subtract the interference from the remaining bits of the larger message, and the second term is the minimum power consumption at $S_1$ with interference from the remaining bits. The maximum then determines the power consumption required to decode the network-coded message at both source nodes.

In the case of $g_{r1}(n)<g_{r2}(n)$, $S_2$ can obtain all messages due to its stronger channel, and $S_1$ has to decode both the network coded bits and the remaining bits from $S_2$ due the assumption that $\lambda_1 \le \lambda_2$. Hence, the power consumption is only determined by $g_{r1}(n)$ since $2^{R^{\mathrm{\mathrm{DNC}}}_{21}(n)}/g_{r1}(n) \ge 1/g_{r2}(n)$. Note that $P^{\mathrm{\mathrm{DNC}}}_{22}(n)$ is determined by the second term in the brackets in (\ref{eq:dnc_uplink_1}).



Recall that the maximum of a series of convex functions of an independent variable is still a convex function with respect to the variable. Hence, it is observed that $P^{\mathrm{\mathrm{DNC}}}_2(n)=P^{\mathrm{\mathrm{DNC}}}_{21}(n)+P^{\mathrm{\mathrm{DNC}}}_{22}(n)$ is a convex function with respect to $R^{\mathrm{\mathrm{DNC}}}_{21}(n)$ and $R^{\mathrm{\mathrm{DNC}}}_{22}(n)$. Interestingly from the above analysis, in the case of $g_{r1}(n)<g_{r2}(n)$, we have $P^{\mathrm{\mathrm{DNC}}}_2(n)=\left(2^{R^{\mathrm{\mathrm{DNC}}}_{21}(n)+R^{\mathrm{\mathrm{DNC}}}_{22}(n)}-1\right)/g_{r1}(n)$.

Further, to better utilize the time resources, a fraction of time resource is assigned to the uplink and the downlink transmission and denoted by $f^{\mathrm{\mathrm{DNC}}}_{i}$ ($i=1$ for the uplink and $i=2$ for the downlink), respectively. We are motivated to minimize the total power consumption by properly assigning power to each sub-carrier and time fractions between the uplink and the downlink of each sub-carrier. Thus, the minimal-power-usage optimization problem over $N$ orthogonal subcarriers, denoted by {\bf P1}, can be formulated as follows.
\begin{align}
\min_{f^{\mathrm{\mathrm{DNC}}}_{i},R^{\mathrm{\mathrm{DNC}}}_{ij}(n)} \quad & \sum_{n=1}^N \sum_{i=1}^2 f^{\mathrm{\mathrm{DNC}}}_iP^{\mathrm{\mathrm{DNC}}}_i(n) \nonumber\\
&+f^{\mathrm{\mathrm{DNC}}}_1P_{rev}+2f^{\mathrm{\mathrm{DNC}}}_2P_{rev}
\label{aim_1}
\end{align}
subject to the rate requirement constraints,
\begin{align}
&\lambda_{i} \le  \sum_{n=1}^N f^{\mathrm{\mathrm{DNC}}}_{1}R^{\mathrm{\mathrm{DNC}}}_{1i}(n) \label{opt_1} \\
&\lambda_{2}-\lambda_1 \le \sum_{n=1}^N f^{\mathrm{\mathrm{DNC}}}_{2}R^{\mathrm{\mathrm{DNC}}}_{21}(n)
      \label{opt_4} \\
&\lambda_{1} \le \sum_{n=1}^N f^{\mathrm{\mathrm{DNC}}}_{2}R^{\mathrm{\mathrm{DNC}}}_{22}(n)    \label{opt_2}
      \end{align}
      and the physical constraints
      \begin{align}
&\sum_{j=1}^2 f^{\mathrm{\mathrm{DNC}}}_{j} \leq 1,  \label{opt_5}\\
&P^{\mathrm{\mathrm{DNC}}}_{11}(n),P^{\mathrm{\mathrm{DNC}}}_{12}(n),P^{\mathrm{\mathrm{DNC}}}_2(n) \leq P_{peak}\label{opt_6}
\end{align}
where in (\ref{aim_1}),
$f^{\mathrm{\mathrm{DNC}}}_1P_{rev}$ accounts for the receiver power consumption at the relay node on the uplink over all subcarriers and $2f^{\mathrm{\mathrm{DNC}}}_2P_{rev}$ is for that consumed by both source nodes on the downlink over all subcarriers.

Note that it can be readily observed that {\bf P1} is not a convex optimization problem due to the product of time fractions $f^{\mathrm{\mathrm{DNC}}}_i$ and transmit rate/power per sub-carrier. For the convexization, we introduce some parameters, i.e., $T^{\mathrm{\mathrm{DNC}}}_{ij}(n) = f^{\mathrm{\mathrm{DNC}}}_iR^{\mathrm{\mathrm{DNC}}}_{ij}(n)$ and $\Theta^{\mathrm{\mathrm{DNC}}}_{ij}(n) = f^{\mathrm{\mathrm{DNC}}}_iP^{\mathrm{\mathrm{DNC}}}_{ij}(n)$. Based on the new variables,
{\bf P1} can be transformed to be {\bf P1'}, as shown below.
\begin{align}
\min_{f^{\mathrm{\mathrm{DNC}}}_{i},T^{\mathrm{\mathrm{DNC}}}_{ij}(n)} \quad & \sum_{n=1}^N \sum_{i=1}^2 \Theta^{\mathrm{\mathrm{DNC}}}_i(n)
+f^{\mathrm{\mathrm{DNC}}}_1P_{rev}
\nonumber\\
& +2f^{\mathrm{\mathrm{DNC}}}_2P_{rev}
\label{aim_1_transformed}
\end{align}
subject to the rate requirement constraints,
\begin{align}
&\lambda_{i} \le  \sum_{n=1}^N T^{\mathrm{\mathrm{DNC}}}_{1i}(n) \label{opt_1_transformed} \\
&\lambda_{2}-\lambda_1 \le \sum_{n=1}^N T^{\mathrm{\mathrm{DNC}}}_{21}(n)
     \label{opt_4_transformed} \\
&\lambda_{1} \le \sum_{n=1}^N T^{\mathrm{\mathrm{DNC}}}_{22}(n)    \label{opt_2_transformed}
      \end{align}
and the time-splitting physical constraint in (\ref{opt_5}) and the peak power constraint in (\ref{opt_6}).

Since $\Theta^{\mathrm{\mathrm{DNC}}}_{ij}(n)$ is the perspective of the convex function $P^{\mathrm{\mathrm{DNC}}}_{ij}(n)$, it can be observed that {\bf P1'} is a convex optimization problem and hence can be solved efficiently by the Karush-Kuhn-Tucker (KKT).

%
To present the optimal solution, we then define $\beta^{\mathrm{\mathrm{DNC}}}_{1i}$ as the multiplier associated with (\ref{opt_1_transformed}), the uplink rate requirement on either side.
$\beta^{\mathrm{\mathrm{DNC}}}_{21}$ accounts for the multiplier corresponding to (\ref{opt_2_transformed}), the downlink rate requirement of the remaining bits.
$\beta^{\mathrm{\mathrm{DNC}}}_{22}$ accounts for the multiplier corresponding to (\ref{opt_2_transformed}), the downlink network-coded message rate requirement. In addition, $\gamma^{\mathrm{\mathrm{DNC}}}$ accounts for the multiplier with the time splitting constraint and $\omega^{\mathrm{\mathrm{DNC}}}_i$ ($i=1,2,r$) is for the peak power constraint at node $i$.
For brevity, the associated Lagrangian function of {\bf P1'} as well as the Karush-Kuhn-Tucker (KKT) conditions are omitted and only the solution to {\bf P1}
is presented in the following lemmas.

\begin{lemma}\label{uplink}
If $g_{1r}(n)>g_{2r}(n)$, the power allocation for the multi-access uplink transmission can be categorized as follows.
\begin{enumerate}
\item if $\beta_{12}^{\mathrm{\mathrm{DNC^*}}} \leq \beta_{11}^{\mathrm{\mathrm{DNC}^*}}$, then
\begin{align}
&P_{11}^{\mathrm{\mathrm{DNC^*}}}(n)=\min\left(P_{peak}, \Delta_{P_{11}^{\mathrm{\mathrm{DNC^*}}}}(n)\right) \nonumber \\
&P_{12}^{\mathrm{\mathrm{DNC^*}}}(n)=0
\end{align}
where $$ \Delta_{P_{11}^{\mathrm{\mathrm{DNC^*}}}}(n)=\left(\beta_{11}^{\mathrm{\mathrm{DNC^*}}}\log_2e-\frac{1}{g_{1r}(n)}\right)^+.$$
\item if $\beta_{12}^{\mathrm{\mathrm{DNC^*}}} > \beta_{11}^{\mathrm{\mathrm{DNC^*}}}$, then

Case I: if $\beta_{11}^{\mathrm{\mathrm{DNC^*}}}g_{11}(n) \leq \beta_{12}^{\mathrm{\mathrm{DNC^*}}}g_{2r}(n)$, then
\begin{align}
\left\{
\begin{array}{ll}
P_{11}^{\mathrm{\mathrm{DNC^*}}}(n)=0\\
P_{12}^{\mathrm{\mathrm{DNC^*}}}(n)=\min\left(P_{peak},\Delta_{P_{11}^{\mathrm{\mathrm{DNC^*}}}}(n)\right)\\
\end{array}
\right.
\end{align}
where $$\Delta_{P_{12}^{\mathrm{\mathrm{DNC^*}}}}(n)=\left(\beta_{12}^{\mathrm{\mathrm{DNC^*}}}\log_2e-\frac{1}{g_{2r}(n)}\right)^+.$$
Case II: if $\beta_{11}^{\mathrm{\mathrm{DNC^*}}}g_{1r}(n)>\beta_{12}^{\mathrm{\mathrm{DNC^*}}}g_{2r}(n)$ and $\beta_{12}^{\mathrm{\mathrm{DNC^*}}}-\beta_{11}^{\mathrm{\mathrm{DNC^*}}} \leq \frac{g_{1r}(n)-g_{2r}(n)}{g_{1r}(n)g_{2r}(n)}$, then
\begin{align}
\left\{
\begin{array}{ll}
P_{11}^{\mathrm{\mathrm{DNC^*}}}(n)=\min\left(P_{peak},\Delta_{P_{11}^{\mathrm{\mathrm{DNC^*}}}}(n)\right)\\
P_{12}^{\mathrm{\mathrm{DNC^*}}}(n)=0\\
\end{array}
\right.
\end{align}
where $$\Delta_{P_{11}^{\mathrm{\mathrm{DNC^*}}}}(n)=\left(\beta_{11}^*\log_2e-\frac{1}{g_{1r}(n)}\right)^+.$$

Case III:if $\beta_{11}^{\mathrm{\mathrm{DNC^*}}}g_{1r}(n)>\beta_{12}^{\mathrm{\mathrm{DNC^*}}}g_{2r}(n)$ and $\beta_{12}^{\mathrm{\mathrm{DNC^*}}}-\beta_{11}^{\mathrm{\mathrm{DNC^*}}} > \frac{g_{1r}(n)-g_{2r}(n)}{g_{1r}(n)g_{2r}(n)}$, then
\begin{align}
\left\{
\begin{array}{llll}
P_{11}^{\mathrm{\mathrm{DNC^*}}}(n)\\
=\min\left(P_{peak}, \Delta_{P_{11}^{\mathrm{\mathrm{DNC^*}}}}(n) \right)\\
P_{12}^{\mathrm{\mathrm{DNC^*}}}(n)\\
=\min\left(P_{peak}, \Delta_{P_{12}^{\mathrm{\mathrm{DNC^*}}}}(n) \right)\\
\end{array}
\right.
\end{align}
where $\Delta_{P_{11}^{\mathrm{\mathrm{DNC^*}}}}(n)$ and $\Delta_{P_{12}^{\mathrm{\mathrm{DNC^*}}}}(n)$ are given in
(\ref{eq:P1_power_1}) and (\ref{eq:P1_power_2}) on top of next page.
\begin{figure*}
\centering
\begin{align}
&\Delta_{P_{11}^{\mathrm{\mathrm{DNC^*}}}}(n)=\left(\frac{(\beta_{11}^{\mathrm{\mathrm{DNC^*}}}g_{1r}(n)-\beta_{12}^{\mathrm{\mathrm{DNC^*}}}g_{2r}(n))\log_2e}{g_{1r}(n)-g_{2r}(n)}\right)^+, \label{eq:P1_power_1}\\
&\Delta_{P_{12}^{\mathrm{\mathrm{DNC^*}}}}(n)=\left(\frac{(\beta_{12}^{\mathrm{\mathrm{DNC^*}}}-\beta_{11}^{\mathrm{\mathrm{DNC^*}}})g_{1r}(n)\log_2e}{g_{1r}(n)-g_{2r}(n)}-\frac{1}{g_{2r}(n)}\right)^+.
\label{eq:P1_power_2}
\end{align}
\end{figure*}
\end{enumerate}
\end{lemma}
Note that $\left(\{\cdot\}\right)^+$ returns $\{\cdot\}$ if it is positive and zero otherwise, and the asterisk denotes optimality. The result of the case with $g_{1r}(n)<g_{2r}(n)$ is similar and omitted here.

For the downlink, the optimal power allocation over each carrier is summarized in Lemma \ref{downlink}.
\begin{lemma}\label{downlink}
The power allocation in the downlink can be categorized as follows.

Case I: If $\beta^{\mathrm{\mathrm{DNC^*}}}_{21}(n) = \beta^{\mathrm{\mathrm{DNC^*}}}_{22}(n)$, the power allocated for the downlink is given by
\begin{align}
P_{2}^{\mathrm{\mathrm{DNC^*}}}(n)=\min\left(P_{peak},\Delta_{P_2^{\mathrm{\mathrm{DNC^*}}}}(n)\right).
\end{align}
where
$$\Delta_{P_2^{\mathrm{\mathrm{DNC^*}}}}(n)=\left(\beta_{21}^{\mathrm{\mathrm{DNC^*}}}\log_2e-\frac{1}{g_{r1}(n)}\right)^+.$$
and the splitting of power allocated for $R^{\mathrm{\mathrm{DNC^*}}}_{21}(n)$ and $R^{\mathrm{\mathrm{DNC^*}}}_{22}(n)$ is subject to the rate requirements in (\ref{opt_4}) and (\ref{opt_2}).

Case II:
Otherwise ($\beta^{\mathrm{\mathrm{DNC^*}}}_{21}(n) \neq \beta^{\mathrm{\mathrm{DNC^*}}}_{22}(n)$), the power allocation for the downlink transmission can be categorized as follows.
\begin{align}
\left\{
\begin{array}{ll}
P_{21}^{\mathrm{\mathrm{DNC^*}}}(n)=\left(\beta_{21}^{\mathrm{\mathrm{DNC^*}}}\log_2e-\frac{1}{g_{r1}(n)}\right)^+\\
P_{22}^{\mathrm{\mathrm{DNC^*}}}(n)=\left(\beta_{22}^{\mathrm{\mathrm{DNC^*}}}\log_2e-\frac{1}{g_{r2}(n)}\right)^+\\
\end{array}
\right.
\end{align}
and $$P_2^{\mathrm{\mathrm{DNC^*}}}(n)=\min\left(P_{peak},P_{21}^{\mathrm{\mathrm{DNC^*}}}(n)+P_{22}^{\mathrm{\mathrm{DNC^*}}}(n)\right).$$
\end{lemma}
Interestingly, with $\beta^{\mathrm{\mathrm{DNC^*}}}_{21}=\beta^{\mathrm{\mathrm{DNC^*}}}_{22}$, we have $P^{\mathrm{\mathrm{DNC^*}}}_2(n)=(2^{R^{\mathrm{\mathrm{DNC^*}}}_{21}(n)+R^{\mathrm{\mathrm{DNC^*}}}_{22}(n)}-1)/g_{r1}$ for all subcarriers in the case of $g_{r1}(n)<g_{r2}(n)$ over all carriers.

\begin{remark}
It is observed that the optimal power allocations at all nodes are in the water-filling structure with a ceiling level that is determined by the peak power constraints at all nodes.
\end{remark}

Further, we present the optimal KKT conditions associated with the optimal time splitting in the case of $g_{r1}(n)<g_{r2}(n)$ and $g_{1r}(n)<g_{2r}(n)$, which are given on top of next page in (\ref{eq:f_1_opt}) and (\ref{eq:f_2_opt}),
\begin{figure*}[!t]
\centering
\begin{align}
&\frac{2^{\frac{T^{\mathrm{\mathrm{DNC^*}}}_{11}(n)+T^{\mathrm{\mathrm{DNC^*}}}_{12}(n)}{f^{\mathrm{\mathrm{DNC^*}}}_1}}}{g_{2r}(n)}
+\left( \frac{1}{g_{1r}(n)}-\frac{1}{g_{2r}(n)}\right)2^{\frac{T^{\mathrm{\mathrm{DNC^*}}}_{11}(n)}{f^{\mathrm{\mathrm{DNC^*}}}_1}}
-\frac{1}{g_{1r}(n)}
+P_{rev}
+\gamma^{\mathrm{\mathrm{DNC^*}}} \nonumber\\
&-\frac{\left(T^{\mathrm{\mathrm{DNC^*}}}_{11}(n)+T^{\mathrm{\mathrm{DNC^*}}}_{12}(n)\right)\ln2}{f^{\mathrm{\mathrm{DNC^*}}}_1} \frac{2^{\frac{T^{\mathrm{\mathrm{DNC^*}}}_{11}(n)+T^{\mathrm{\mathrm{DNC^*}}}_{12}(n)}{f^{\mathrm{\mathrm{DNC^*}}}_1}}}{g_{2r}(n)}
-\left( \frac{1}{g_{1r}(n)}-\frac{1}{g_{2r}(n)}\right)2^{\frac{T^{\mathrm{\mathrm{DNC^*}}}_{11}(n)}{f^{\mathrm{\mathrm{DNC^*}}}_1}}
\frac{T^{\mathrm{\mathrm{DNC^*}}}_{11}(n)\ln 2}{f^{\mathrm{\mathrm{DNC^*}}}_1}
  \nonumber\\
&+\omega^{\mathrm{\mathrm{DNC^*}}}_1\left( \frac{2^{\frac{T^{\mathrm{\mathrm{DNC^*}}}_{11}(n)}{f^{\mathrm{\mathrm{DNC^*}}}_1}}-1}{g_{1r}(n)}
- \frac{2^{\frac{T^{\mathrm{\mathrm{DNC^*}}}_{11}(n)}{f^{\mathrm{\mathrm{DNC^*}}}_1}}}{g_{1r}(n)} \frac{T^{\mathrm{\mathrm{DNC^*}}}_{11}(n)\ln2}{f^{\mathrm{\mathrm{DNC^*}}}_1}      -P_{peak}
\right) \nonumber\\
&+\omega^{\mathrm{\mathrm{DNC^*}}}_2\Biggl( \frac{2^{\frac{T^{\mathrm{\mathrm{DNC^*}}}_{11}(n)}{f^{\mathrm{\mathrm{DNC^*}}}_1}}\left( 2^{\frac{T^{\mathrm{\mathrm{DNC^*}}}_{12}(n)}{f^{\mathrm{\mathrm{DNC^*}}}_1}} -1 \right)}{g_{2r}(n)} \nonumber \\
&- \left( \frac{2^{\frac{T^{\mathrm{\mathrm{DNC^*}}}_{11}(n)+T^{\mathrm{\mathrm{DNC^*}}}_{12}(n)}{f^{\mathrm{\mathrm{DNC^*}}}_1}}}{g_{2r}(n)}\right)
\frac{\left(T^{\mathrm{\mathrm{DNC^*}}}_{11}(n)+T^{\mathrm{\mathrm{DNC^*}}}_{12}(n)\right)\ln2}{f^{\mathrm{\mathrm{DNC^*}}}_1}
-\left( \frac{2^{\frac{T^{\mathrm{\mathrm{DNC^*}}}_{11}(n)}{f^{\mathrm{\mathrm{DNC^*}}}_1}}}{g_{2r}(n)}\right)\frac{T^{\mathrm{\mathrm{DNC^*}}}_{11}(n)\ln2}{f^{\mathrm{\mathrm{DNC^*}}}_1}
-P_{peak}
\Biggr)\nonumber \\
=&0   \label{eq:f_1_opt}\\
&\frac{2^{\frac{T^{\mathrm{\mathrm{DNC^*}}}_{21}(n)+T^{\mathrm{\mathrm{DNC^*}}}_{22}(n)}{f^{\mathrm{\mathrm{DNC^*}}}_2}}}{g_{r1}(n)}
-\frac{\left(T^{\mathrm{\mathrm{DNC^*}}}_{21}(n)+T^{\mathrm{\mathrm{DNC^*}}}_{22}(n)\right)\ln2}{f^{\mathrm{\mathrm{DNC^*}}}_2} \frac{2^{\frac{T^{\mathrm{\mathrm{DNC^*}}}_{21}(n)+T^{\mathrm{\mathrm{DNC^*}}}_{22}(n)}{f^{\mathrm{\mathrm{DNC^*}}}_2}}}{g_{r1}(n)}
+2P_{rev}
+\gamma^{\mathrm{\mathrm{DNC^*}}} \nonumber\\
&+\omega^{\mathrm{\mathrm{DNC^*}}}_r\left( \frac{2^{\frac{T^{\mathrm{\mathrm{DNC^*}}}_{21}(n)+T^{\mathrm{\mathrm{DNC^*}}}_{22}(n)}{f^{\mathrm{\mathrm{DNC^*}}}_2}}}{g_{r1}(n)}
-\frac{\left(T^{\mathrm{\mathrm{DNC^*}}}_{21}(n)+T_{22}(n)\right)\ln2}{f^{\mathrm{\mathrm{DNC^*}}}_2} \times \frac{2^{\frac{T^{\mathrm{\mathrm{DNC^*}}}_{21}(n)+T^{\mathrm{\mathrm{DNC^*}}}_{22}(n)}{f^{\mathrm{\mathrm{DNC^*}}}_2}}}{g_{r1}(n)}-P_{peak}
\right)\nonumber \\
=&0. \label{eq:f_2_opt}
\end{align}
\end{figure*}
where $f^{\mathrm{\mathrm{DNC^*}}}_i$ and $T^{\mathrm{\mathrm{DNC^*}}}_{ij}(n)$
are optimization variables in {\bf P1'}. The equations for the cases of other channel gains can be similarly derived and are omitted here.
Note also that (\ref{eq:f_1_opt}) and (\ref{eq:f_2_opt}) are transcendental equations and $f_i^{\mathrm{\mathrm{DNC^*}}}$ ($i=1,2$) can only be solved numerically.

\section{PNC}

A PNC scheme was discussed in \cite{wilson2010joint} for symmetric traffic,
symmetric channel TWRNs, where the lattice coding is employed to achieve the rate region
For Gaussian channels. Our work extends the model of \cite{wilson2010joint}
to the scenario of asymmetric traffic and asymmetric channel
based on the PNC scheme with a triple transmission mode,
where two modes are for uplink and the other is for downlink.

In the first PNC uplink mode, each source node simultaneously transmits an equal amount of bits on each carrier to $R$, and $R$ simply decodes a function message of them, i.e., decode $\widehat{a+b_1}$ if $S_1$ transmits $a$ and $S_2$ transmits $b_1$ on a carrier, where $P_{11}^{\mathrm{PNC}}(n)g_{1r}(n)=P_{12}^{\mathrm{PNC}}(n)g_{2r}(n)$ is required to make the SNR of each received message equal. This is to avoid any additional interference when decoding the function message. Note that in $P_{1i}^{\mathrm{PNC}}(n)$ ($i=1,2$) the first subscript $1$ denotes mode index and the second subscript $i$ denotes the labeling of the source nodes.
Hence, from \cite{wilson2010joint}, the achievable transmit rate with PNC at carrier $n$ is given by,
\begin{align}
R_1^{\mathrm{PNC}}(n)
=&\log_2 \left( \frac{1}{2} + P_{11}^{\mathrm{PNC}}(n)g_{1r}(n)  \right) \nonumber\\
=&\log_2 \left( \frac{1}{2} + P_{12}^{\mathrm{PNC}}(n)g_{2r}(n)  \right). \label{eq:PNC_1}
\end{align}
The transmit power of $S_i$ in mode 1 at carrier $n$ is then given by,
\begin{align}
P_{1i}^{\mathrm{PNC}}(n)=
\frac{2^{R_1^{\mathrm{PNC}}(n)}-\frac{1}{2}}{g_{ir}(n)}.
\end{align}
The sum power at carrier $n$ in mode 1 hence is $P_{1}^{\mathrm{PNC}}(n)=\sum_{i=1}^2 P_{1i}^{\mathrm{PNC}}(n)$.

In the second mode, $S_2$ transmits the remaining bits of the larger message to $R$ (i.e., $b_2$) given the assumption $\lambda_1<\lambda_2$. The transmit rate at carrier $n$ of this mode is given by,
\begin{align}
R_2^{\mathrm{PNC}}(n)&=\log_2 \left( 1 + P_2^{\mathrm{PNC}}(n)g_{2r}(n) \right).
\end{align}

Transmission of the third mode of PNC is identical to that of the downlink of DNC and the details are omitted for brevity.
Note that $R_{32}^{\mathrm{PNC}}(n)$ is for network-coded message and $R_{31}^{\mathrm{PNC}}(n)$ is for the remaining bits of the larger message over carrier $n$. The power required for transmission of $R_{32}^{\mathrm{PNC}}(n)$ and $R_{31}^{\mathrm{PNC}}(n)$ at the $n$th subcarrier is denoted by $P_3^{\mathrm{PNC}}(n)$

Let $f_i^{\mathrm{PNC}}(n)$ ($i=1,2,3$) denote the time resource allotted to the $i$th mode of PNC over the $n$th carrier. We can formulate a problem employing PNC to minimize the total power consumption of the entire system. It is referred to as {\bf P2} and is formulated as follows.
\begin{align}
\min_{f_{i}^{\mathrm{PNC}},R_{ij}^{\mathrm{PNC}}(n)} \quad & \sum_{n=1}^N \sum_{i=1}^3 f_i^{\mathrm{PNC}}P_i^{\mathrm{PNC}}(n)
+2f_3^{\mathrm{PNC}}P_{rev} \nonumber\\
&+(f_1^{\mathrm{PNC}}+f_2^{\mathrm{PNC}})P_{rev}
\label{aim_2}
\end{align}
subject to the constraints as follows,
\begin{align}
&\lambda_{1} \le \sum_{n=1}^N f_{1}^{\mathrm{PNC}}R_{1}^{\mathrm{PNC}}(n) \label{opt_21}\\
&\lambda_{1} \le \sum_{n=1}^N f_{3}^{\mathrm{PNC}}R_{32}^{\mathrm{PNC}}(n) \label{opt_22}\\
&\lambda_{2} \le  \sum_{n=1}^N \sum_{i=1}^2f_{i}^{\mathrm{PNC}}R_{i}^{\mathrm{PNC}}(n)  \label{opt_23} \\
&\lambda_{2}-\lambda_1 \le \sum_{n=1}^N f_{3}^{\mathrm{PNC}}R_{31}^{\mathrm{PNC}}(n) \label{opt_24} \\
&\sum_{j=1}^3 f_{j}^{\mathrm{PNC}} \leq 1,    \label{opt_25}\\
&P_{11}^{\mathrm{PNC}}(n),P_{12}^{\mathrm{PNC}}(n),P_2^{\mathrm{PNC}}(n),
P_3^{\mathrm{PNC}}(n) \leq P_{peak} \label{opt_26}
\end{align}
where $\left(f_1^{\mathrm{PNC}}+f_2^{\mathrm{PNC}}\right)P_{rev}$ accounts for the receive power consumption at the relay node over all subcarriers, and $2f_3^{\mathrm{PNC}}P_{rev}$ is the receive power consumption at both source nodes over all subcarriers.

Note that {\bf P2} is not a convex optimization problem due to the quadratic terms. However, by employing a similar procedure as in {\bf P1}, {\bf P2} can be transformed to be an equivalent convex optimization problem {\bf P2'} and is solved efficiently via KKT conditions. The detailed analysis is however omitted.

Further, the optimal power allocation on downlink of PNC is identical to that of {\bf P1} and hence omitted for brevity. We only focus on the optimal power allocation to the first two PNC modes on uplink, which is listed as follows.
\begin{align}
&P_{1}^{\mathrm{PNC}}(g_{1r}(n),g_{2r}(n))
=\min\left(2P_{peak}, \Delta_{P_1^{\mathrm{PNC}^*}}\right) \label{eqn:pncopt1}\\
&P_{1i}^{\mathrm{PNC}}(g_{1r}(n),g_{2r}(n))=\min\left(P_{peak}, \Delta_{P_{1i}^{\mathrm{PNC}^*}} \right) \label{eqn:pncopt2}\\
&P_2^{\mathrm{PNC}}(g_{r1}(n),g_{r2}(n))\nonumber\\ =&\min\left(P_{peak},\left(\beta_2^{\mathrm{PNC}^*}\log_2 e - \frac{1}{g_{2r}(n)}\right)^+\right) \label{eqn:pncopt3}
\end{align}
where
$$\Delta_{P_1^{\mathrm{PNC}^*}} = \left(\beta_1^{\mathrm{PNC}^*} \log_2 e - \frac{1}{2}\left( \frac{1}{g_{1r}(n)}+\frac{1}{g_{2r}(n)} \right) \right)^+$$
$$\Delta_{P_{1i}^{\mathrm{PNC}^*}} = \left( \frac{\beta_1^{\mathrm{PNC}^*} \log_2 eg_{3-i,r}(n)}{g_{1r}(n)+g_{2r}(n)}-\frac{1}{2g_{ir}(n)}\right)^+$$
where $2P_{peak}$ in (\ref{eqn:pncopt1}) is the sum peak power constraint of the two source nodes, and $\beta_i^{\mathrm{PNC}^*}$ ($i=1,2$) is the optimal multiplier
with respect to the corresponding rate requirements.
In addition, the optimal KKT conditions for $f_i^{\mathrm{PNC}}$ can be similarly derived and omitted for brevity.


\section{Hybrid PNC/DNC Switching Scheme}
Each PNC and DNC can individually operate for achieving optimal power efficiency, mostly under different SNR region (which can be clearly observed in the numerical analysis results). Thus, we are motivated to develop a hybrid PNC/DNC switching scheme that adaptively operates in either PNC or DNC for each message transmission according the channel SNR, in order to further enhance the performance.

To analyze the hybrid scheme, we firstly employ an indication variable ${\mathrm{I}}_{\mathrm{PNC}}(n)$ for PNC and an indication variable ${\mathrm{I}}_{\mathrm{DNC}}(n)$ for DNC, respectively, where ${\mathrm{I}}_{\mathrm{scheme}}(n)=1$ ($\mathrm{scheme} \in \{{\mathrm{PNC}},{\mathrm{DNC}}\}$) if on subcarrier $n$ the specified scheme is employed and zero otherwise.
Hence, similar to the formulation of {\bf P1} and {\bf P2}, we can formulate this hybrid scheme, namely {\bf Popt}, in (\ref{eq:obj_P3}) on top of next page.
\begin{figure*}
\centering
\begin{align}
\min_{f_{i}^{\mathrm{PNC}},f_{i}^{\mathrm{DNC}},
R_{ij}^{\mathrm{PNC}}(n),R_{ij}^{\mathrm{DNC}}(n),{\mathrm{I}}_{\mathrm{PNC}}(n),
{\mathrm{I}}_{\mathrm{DNC}}(n)}
\quad & \sum_{n=1}^N \biggl( \sum_{i=1}^3 {\mathrm{I}}_{\mathrm{PNC}}(n)   f_i^{\mathrm{PNC}}P_i^{\mathrm{PNC}}(n) + \sum_{i=1}^2  {\mathrm{I}}_{\mathrm{DNC}}(n) f_i^{\mathrm{DNC}}P_i^{\mathrm{DNC}}(n)
\biggr) \nonumber\\
&+(f_1^{\mathrm{PNC}}+f_2^{\mathrm{PNC}})P_{rev}+2f_3^{\mathrm{PNC}}P_{rev}
\label{eq:obj_P3}
\end{align}
\end{figure*}
The constraints are as follows. (\ref{opt_P3_1})-(\ref{opt_P3_4}) are for rate requirements of the system. (\ref{opt_P3_5}) and (\ref{opt_P3_6}) are for physical constraints of the time splitting. Note also that (\ref{opt_P3_constraint_t_1}) and (\ref{opt_P3_constraint_t_2}) hold due to the half-duplex assumption that each node can either transmit or receive on all carriers simultaneously, but not both. (\ref{opt_P3_7}) is the peak power constraint at the transmitter. (\ref{opt_P3_8}) is the constraint on the indication variables, which can only take values in the set $\{0,1\}$. Finally, (\ref{opt_P3_9}) and (\ref{opt_P3_10}) indicate that only one scheme can be used on a carrier at any moment.
\begin{figure*}
\centering
\begin{align}
&\lambda_{1} \le \sum_{n=1}^N {\mathrm{I}}_{\mathrm{PNC}}(n) f_{1}^{\mathrm{PNC}}R_{1}^{\mathrm{PNC}}(n)
+\sum_{n=1}^N {\mathrm{I}}_{\mathrm{DNC}}(n) f_{1}^{\mathrm{DNC}}R_{11}^{\mathrm{DNC}}(n)
\label{opt_P3_1}\\
&\lambda_{1} \le \sum_{n=1}^N {\mathrm{I}}_{\mathrm{PNC}}(n) f_{3}^{\mathrm{PNC}}R_{32}^{\mathrm{PNC}}(n)
+\sum_{n=1}^N {\mathrm{I}}_{\mathrm{DNC}}(n) f_{2}^{\mathrm{DNC}}R_{22}^{\mathrm{DNC}}(n)
\label{opt_P3_2}\\
&\lambda_{2} \le  \sum_{n=1}^N \sum_{i=1}^2
{\mathrm{I}}_{\mathrm{PNC}}(n) f_{i}^{\mathrm{PNC}}R_{i}^{\mathrm{PNC}}(n)
+{\mathrm{I}}_{\mathrm{DNC}}(n) f_{1}^{\mathrm{PNC}}R_{12}^{\mathrm{PNC}}(n)
\label{opt_P3_3} \\
&\lambda_{2}-\lambda_1 \le \sum_{n=1}^N {\mathrm{I}}_{\mathrm{PNC}}(n) f_{3}^{\mathrm{PNC}}R_{31}^{\mathrm{PNC}}(n)
+\sum_{n=1}^N{\mathrm{I}}_{\mathrm{DNC}}(n) f_{2}^{\mathrm{DNC}}R_{21}^{\mathrm{DNC}}(n)
 \label{opt_P3_4} \\
&\sum_{j=1}^3 f_{j}^{\mathrm{PNC}} = 1,    \label{opt_P3_5}\\
&\sum_{j=1}^2 f_{j}^{\mathrm{DNC}} = 1,    \label{opt_P3_6}\\
&f_{1}^{\mathrm{PNC}}+f_{2}^{\mathrm{PNC}} = f_{1}^{\mathrm{DNC}}
\label{opt_P3_constraint_t_1}\\
&f_{3}^{\mathrm{PNC}} = f_{2}^{\mathrm{DNC}}
\label{opt_P3_constraint_t_2}\\
&P_{1i}^{\mathrm{PNC}}(n),P_2^{\mathrm{PNC}}(n),
P_3^{\mathrm{PNC}}(n), P_{1i}^{\mathrm{DNC}}(n), P_2^{\mathrm{DNC}}(n) \leq P_{peak}, \quad i=1,2 \label{opt_P3_7}\\
& {\mathrm{I}}_{\mathrm{PNC}}(n) , {\mathrm{I}}_{\mathrm{DNC}}(n)  \in \{0,1\}
\label{opt_P3_8}\\
& {\mathrm{I}}_{\mathrm{PNC}}(n) {\mathrm{I}}_{\mathrm{DNC}}(n)=0 \label{opt_P3_9}\\
& {\mathrm{I}}_{\mathrm{PNC}}(n) +{\mathrm{I}}_{\mathrm{DNC}}(n)=1 \label{opt_P3_10}
\end{align}
\end{figure*}

From above, it can be easily found that {\bf Popt} is not a convex function, and the global optimal solution is hence not tractable. However, by employing similar approach to that in {\bf P1} and {\bf P2} where these indication variables over all subcarriers are specified, {\bf Popt} can be also be convexified. Hence, an iterative algorithm is developed to find a promising sub-optimal solution. The presented algorithm, termed as Algorithm 1, is given as follows.

\begin{enumerate}
\item Input: Rate pair ($\lambda_1$,$\lambda_2$), average channel gains $g_{ij}(n)$ ($i,j=r,1,2$) as well as a predefined termination threshold $\epsilon$.
\item Initialization: Select DNC as the initial scheme if {\bf P1} outperforms {\bf P2} and PNC otherwise, and set the associated indication variables.
\item Iteration: With the specified indication variables, find the global-optimal solution for the degraded problem, i.e., sub-{\bf Popt}, as well as return the allocated time and power, and the rate over each subcarrier.
\item On each subcarrier, with the assigned time and rate, compute the power consumption if the other scheme is selected on this subcarrier.
    If the other scheme consumes less power, it is selected and the associated indication variables are updated,
    otherwise the indication variables remain unchanged.
    Compute the new total power consumption with the assigned time and rate but
    the updated schemes.

\item  Compare the two total power consumption found in Step 3) and Step 4). If the difference is greater than $\epsilon$, go back to Step 3) for another iteration.
    Otherwise, go to Step 6) and terminate.
\item Output: return the found sub-optimal solution.
\end{enumerate}

As observed in Algorithm 1, we iteratively switch to a better scheme on each subcarrier until a predefined threshold is reached. One one hand, the algorithm takes the best of {\bf P1} and {\bf P2} at the beginning. With the given indication variables, we solve the sub-problem of {\bf Popt} in Step 3) that returns a global-optimal solution and hence outperform the previous iterations. On the other hand, we always select the power-efficient scheme over all subcarriers given the assigned time and rate, which definitely improves the sub-optimal solution from Step 3). Thus, we can expect that the iterated solution must be the most power-efficient. This will be attested in the following section.

For convergence, it is readily observed that the found optimal total power usage is lower bounded by zero and decreases in each iteration, our algorithm hence will converge to some certain point.  Note also that we are not able to analytically discuss how fast the algorithm converges, however, in the case studies, we find tens of iterations are sufficient to reach the stopping criterion in Step 5).

It is also worth noting that our iterative algorithm is implemented offline, as the required information, i.e., the rate pair requirement and the channel gains,
\footnote{Note that we assume the channels are Gaussian over the carriers
in this work.}are assumed to be known a priori at the transmitters.  It hence results in a look-up table, with the suggested scheme and the allocated resource per carrier along with the corresponding channel state information and required rates as records. This look-up table is calculated at the central-controller, i.e., the relay node and is shared with the two source nodes. In this sense, the overhead of the proposed algorithm in real-time transmission is therefore guaranteed to be negligible.

\section{Numerical Results}
   \begin{figure}[t]
   \centering
   \includegraphics[width = 12cm]{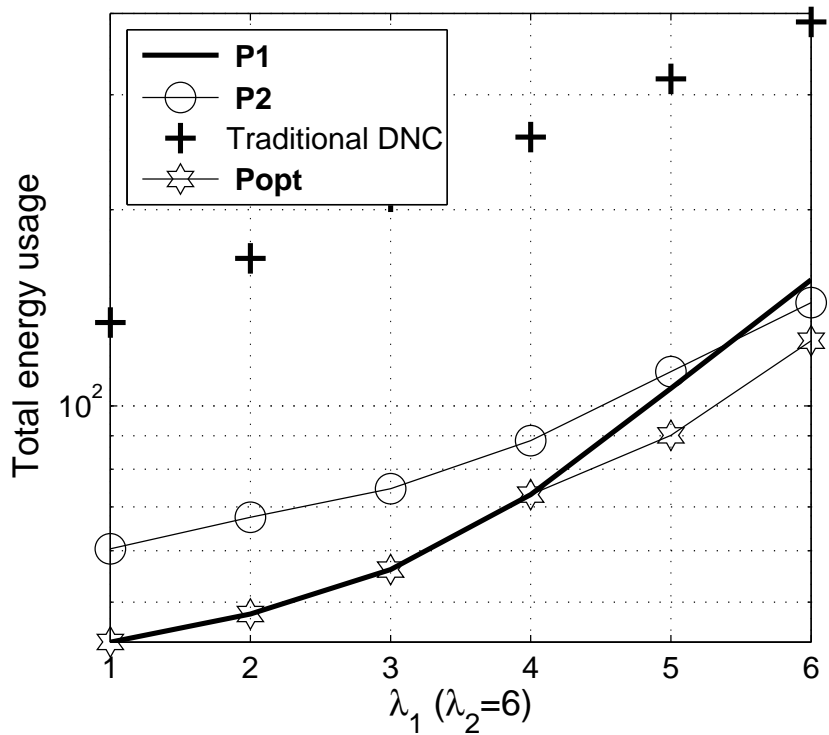}
   \caption{Total power consumption of different schemes for four carriers under small asymmetric traffic scenario.} \label{fig:observation_asymmetric}
   \end{figure}
   \begin{figure}[t]
   \centering
   \includegraphics[width = 12cm]{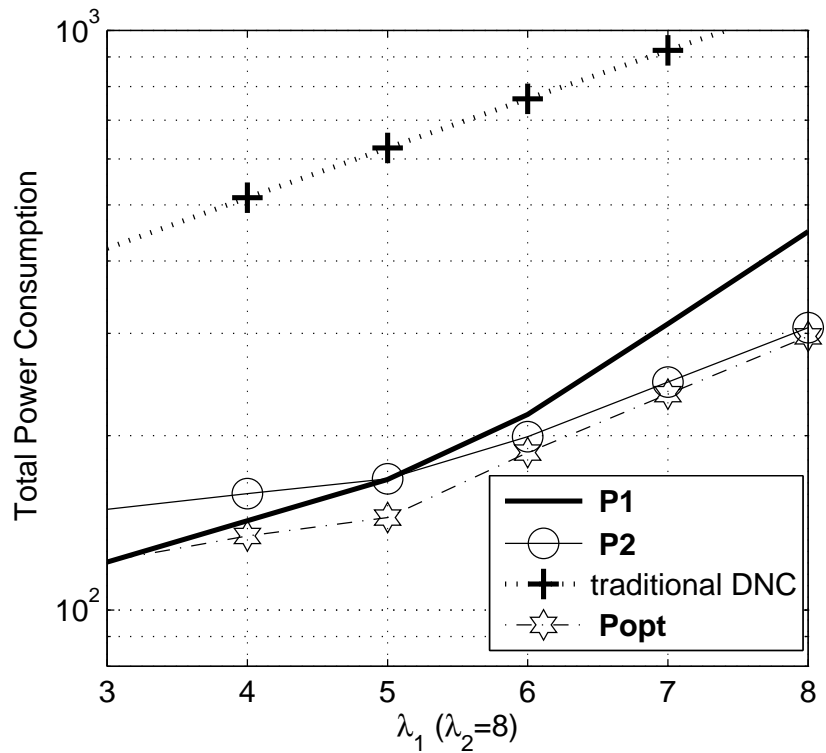}
   \caption{Total power consumption of different schemes for four carriers under relatively large asymmetric traffic scenario.} \label{fig:observation_asymmetric_2}
   \end{figure}
   \begin{figure}[t]
   \centering
   \includegraphics[width = 12cm]{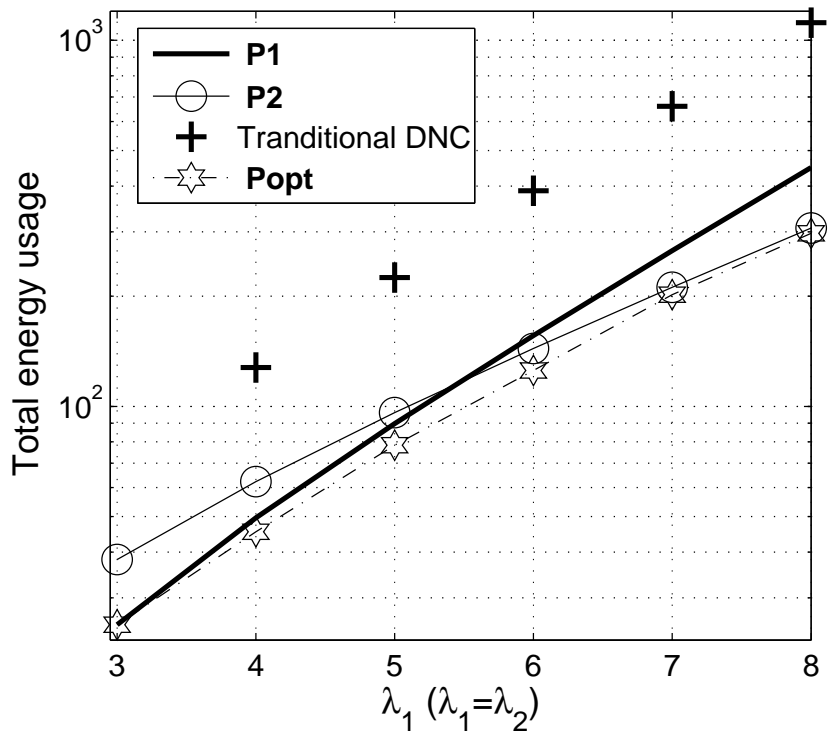}
   \caption{Total power consumption comparison for four carriers under symmetric traffic scenario.} \label{fig:observation_symmetric}
   \end{figure}

   \begin{figure}[t]
   \centering
   \includegraphics[width = 12cm]{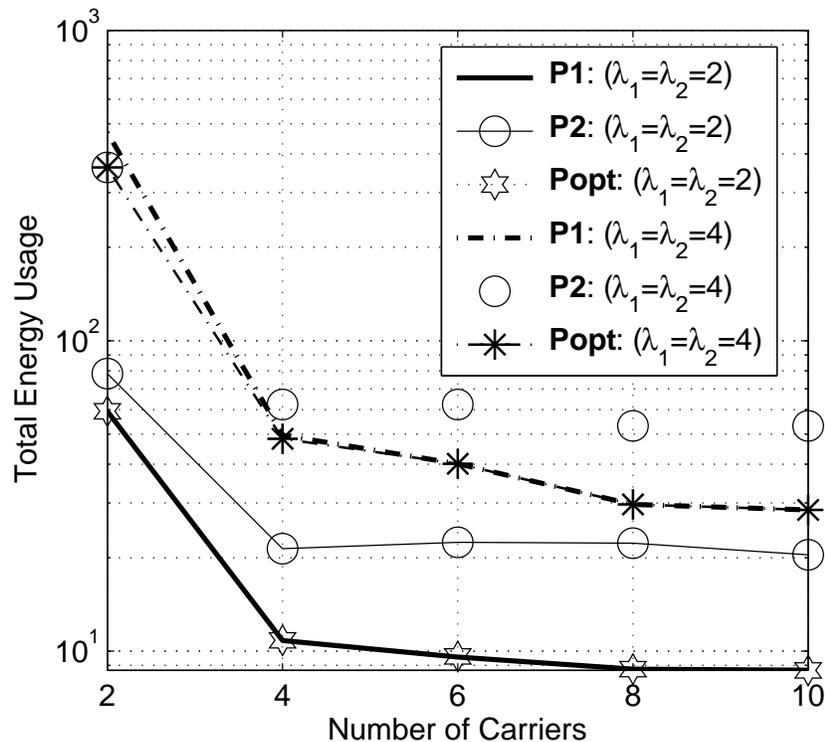}
   \caption{Total power consumption comparison versus carrier number with small traffic scenario.} \label{fig:carrier3}
   \end{figure}

   \begin{figure}[t]
   \centering
   \includegraphics[width = 12cm]{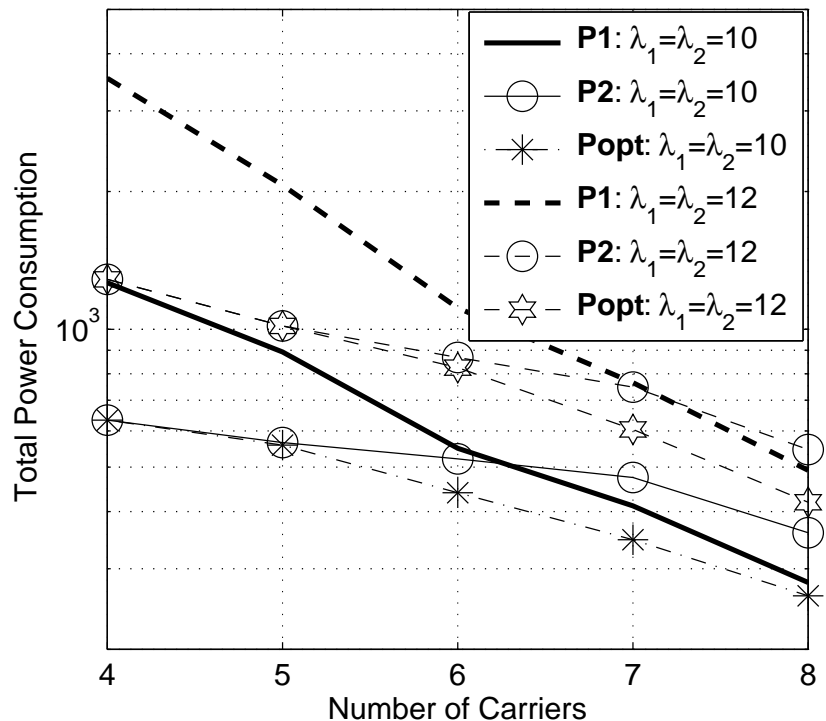}
   \caption{Total power consumption comparison versus carrier number with relatively large traffic scenario.} \label{fig:carrier5}
   \end{figure}

   \begin{figure}[t]
   \centering
   \includegraphics[width = 12cm]{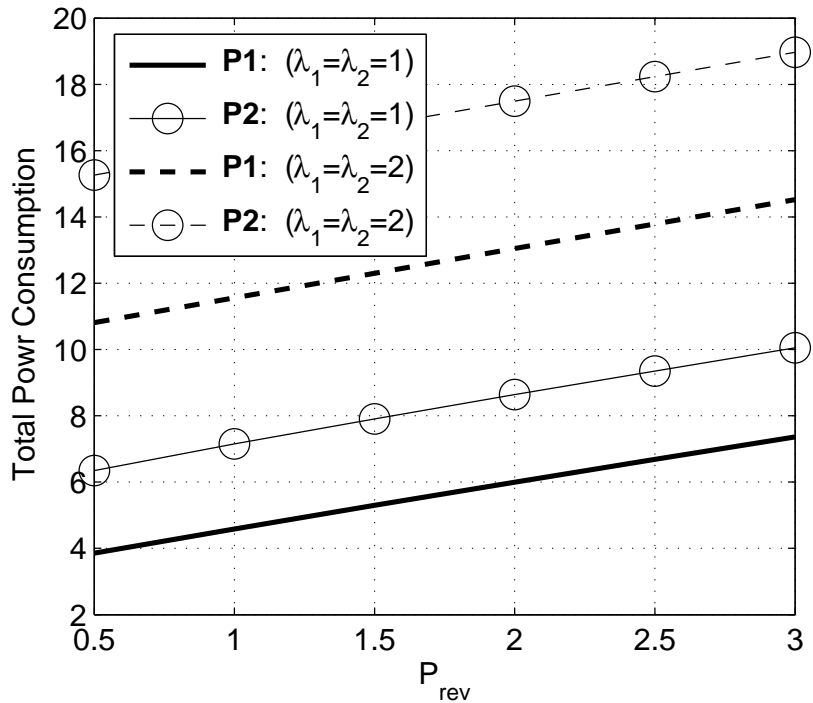}
   \caption{Total power consumption comparison versus receive power.} \label{fig:recive_1}

   \end{figure}
   \begin{figure}[t]
   \centering
   \includegraphics[width = 12cm]{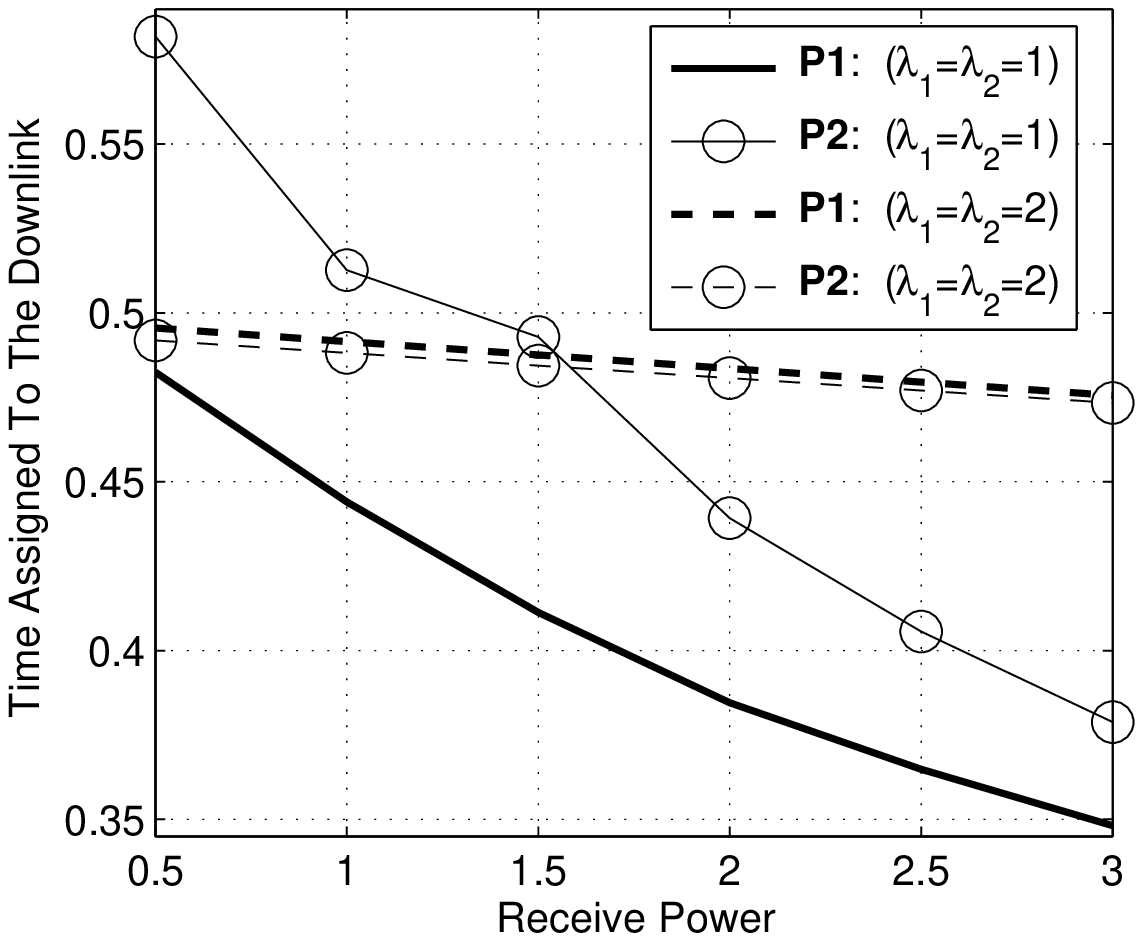}
   \caption{Optimal time assigned to the downlink versus receive power consumption.} \label{fig:recive_2}
   \end{figure}

We now present numerical results to verify our findings.
To reduce computation burden, we assume that there are totally $4$ SCs, and the bandwidth of each SC is assumed to be unity. All channels are assumed to be independent Gaussian channels. Further, we assume channel reciprocity over all links, i.e., $g_{ir}(n)=g_{ri}(n)$. The channel gains of the four carriers, i.e., $g_{1r}(n),g_{2r}(n)$,  in simulation are set to be:
(0.2049, 2.0637),  (0.0989, 0.0906),  (0.4583, 1.2783) and (2.3275, 0.6035), respectively. Noise at each node over each subcarrier is assumed to be Gaussian with zero mean and unit variance.

The peak power constraint is $2000$ J/sec per carrier at each node. The receive power consumption of each node is taken as $0.5$ J/sec, if not noted. For fair comparison, we also compare the considered schemes ({\bf P1} to {\bf P2}) that employ the traditional DNC scheme with time-sharing uplink for source messages
and time-sharing downlink for both network-coded message and residual bits transmission.



In Fig. \ref{fig:observation_asymmetric}, we compare the proposed schemes as well as the traditional three-slot DNC under the asymmetric rate pair requirement. We see that all the proposed schemes significantly outperform the three-slot DNC due to better use of spectrum resources. It is also observed that {\bf P1} achieves better performance than {\bf P2} with $\lambda_1<5.5$ bit/sec and worse than {\bf P2} with $\lambda_1>5.5$ bit/sec. This is due to the fact that the interference of the multi-access uplink dominates and deteriorates the performance of DNC particularly in the high SNR regime. On the other hand, with small data arrival rates or low-SNR regimes, the performance of PNC becomes much outperformed as observed in (\ref{eq:PNC_1}). Further, the proposed hybrid scheme {\bf Popt} is observed to perform the best under all the considered scenarios, thanks to the fact that it always takes the best scheme among all carriers. It is interesting to note that, the performance of {\bf Popt} matches with {\bf P1} with $\lambda_1<4$ bit/sec, as in this small data arrival rate case, DNC performs best over all subcarriers. With $\lambda_1>4$ bit/sec, however, it beats both DNC and PNC, due to the opportunistic selection of the best scheme over all subcarriers.

In Fig. \ref{fig:observation_asymmetric_2}, we consider the case with relatively large traffic from both sides.
As expected, PNC outperforms DNC under high SNR regimes; and {\bf Popt} performs the best under all scenarios. Further, the gap between the hybrid scheme and PNC/DNC shrinks with the increasing/decreasing data rates, respectively. This intuitively follows from the fact that, under very low (very high) SNR scenarios, one scheme is always better than the other; while in the medium SNR scenario, each scheme might be potentially better and hence the opportunism characteristic of {\bf Popt} helps to perform the best.

In Fig. \ref{fig:observation_symmetric}, we compare different network coding schemes under the symmetric rate pair requirement (i.e., $\lambda_1=\lambda_2$). We can see that both {\bf P1} and {\bf P2} significantly outperform the traditional DNC. Further, the performance gap between PNC and DNC increases with the increase of the arrival rates. Therefore, we can conclude when the arrival rate is high, the interference in the multi-access channel of DNC deteriorates its performance and performs worse than PNC; while when  the arrival rate is low, the fact of the uplink rate is somewhat constrained by the minimum channel gain of the uplink channels dominates its performance, which makes DNC performs worse than PNC.
Finally, {\bf Popt} is seen to performs best as expected.

Since both {\bf P1} and {\bf P2} significant outperforms the traditional 3-slot DNC scheme, we shall only consider the two schemes as well as {\bf Popt} in the following numerical analysis, where performance versus the number of carriers as well as the impact of receive power are all considered.

\subsection{Performance Versus Carrier Number}
In Fig. \ref{fig:carrier3} and Fig. \ref{fig:carrier5}, we simulate the total power consumption versus the number of carriers in both relatively small and large data traffic scenarios.
It is seen that the total power consumption is declined with the increasing number of carriers, which meets the intuition. Further, the gap of DNC to PNC increases as the increasing number of carriers with relatively lower arrival rate requirement. This follows from the fact that, the rate allotted to each carrier is smaller with the increasing number of carriers, as is similar in low-rate requirement. However, PNC is still better with very high arrival rate pair in Fig. \ref{fig:carrier5}, as in this case the SNR is relatively high even by averaged over the carriers. Lastly, {\bf Popt} is observed to be always the best among all schemes in terms of power consumption.

\subsection{Performance Versus Receive Power Consumption}
In Fig. \ref{fig:recive_1} and Fig. \ref{fig:recive_2}, the impact of receive power on both total power consumption and time assigned to the downlink is shown.
As expected, with the increasing receive power consumption of the system, the total power consumption increases while the time assigned to the downlink is decreased for both DNC and PNC schemes. This follows that, both source nodes consumes receive power over the downlink, while over the uplink, only the relay node consumes receive power.
We have also seen that with the changed time assigned to both the uplink and the downlink, the rate assigned is also changed, as well as the optimal water level in the transmit power allocations.
Finally, the impact of receive power usage gradually deteriorates with the increasing of rate requirements, i.e., the increasing of transmit power consumption.

\section{Conclusions}
In this work, we considered a three-node TWRN over a multi-carrier system by aiming at minimizing the total power consumption. DNC, PNC, and a novel hybrid DNC/PNC schemes are analyzed and formulated into optimization problems. They were found to be nonconvex optimization problems at first glance but were convexified for better computation tractability, in which their global optimal solutions were obtained by KKT conditions. Extensive simulation was conducted to examine the considered system under the three schemes and compared them to the legacy ones. We found that PNC can achieve better performance than DNC under high rate pair requirements while becomes outperformed with low data rate pair requirements. Such a trend holds for various scenarios considered in the simulation settings. The proposed hybrid PNC/DNC switching scheme, by opportunistically taking the best advantage of the two, is proved to achieve the best performance in all the considered scenarios.

%

\end{document}